\documentclass[journal,a4paper]{IEEEtran}
\usepackage{color}
\usepackage{amsmath,amsfonts}
\usepackage{algorithmic}
\usepackage{array}
\usepackage{textcomp}
\usepackage{stfloats}
\usepackage{url}
\usepackage{verbatim}
\usepackage{subcaption}
\usepackage{graphicx}
\usepackage{cite}
\usepackage{amssymb}

\usepackage[ruled]{algorithm2e}
\usepackage{makecell}
\usepackage{multicol}

\makeatletter
\newcommand{\removelatexerror}{\let\@latex@error\@gobble}
\makeatother
\usepackage[justification=centering]{caption}
\begin{document}

\title{Semantic Communication Meets Heterogeneous Network: Emerging Trends, Opportunities, and Challenges}

\author{Guhan Zheng, Qiang Ni, Aryan Kaushik, Lixia Yang, Yushi Wang, and Charilaos Zarakovitis

\thanks{G. Zheng is with the School of Communication and Information Engineering, Shanghai University, 200444, China (e-mail: gzheng@shu.edu.cn).}

\thanks{Q. Ni and Y. Wang are with the School of Computing and Communications, Lancaster University, LA1 4WA, UK (e-mail: \{q.ni, y.wang216\}@lancaster.ac.uk).}
\thanks{A. Kaushik is with the RakFort, Ireland, and IIITD, India. (email: a.kaushik@ieee.org).}
\thanks{L. Yang is with the School of Electronics and Information, Anhui University, 230601, China. (e-mail: lixiayang@yeah.net).}
\thanks{C. Zarakovitis is with AXON Logic, Greece (Email: c.zarakovitis@axonlogic.gr).}
}

\maketitle
\thispagestyle{empty}

\begin{abstract}
Recent developments in machine learning (ML) techniques enable users to extract, transmit, and reconstruct information semantics at the semantic level through ML-based semantic communication (SemCom). This significantly increases network spectral efficiency and transmission robustness. The semantic codecs among various users and modalities, based on ML, however, inevitably experience semantic drift and necessitate collaborative updating to preserve transmission quality. The various heterogeneous characteristics of most networks, in turn, introduce emerging but unique challenges for semantic codec updating that are different from other general ML model updating. In this article, we propose a heterogeneity-aware semantic codec updating scheme to achieve efficient and reliable updating in heterogeneous networks. We begin with the introduction of the core components of SemCom and then highlight key issues in semantic codec updating under network heterogeneity, discussing several potential methods. Furthermore, the scheme is provided with performance metrics. Future research directions for advancing SemCom in complex, multi-modal environments are also discussed.  
\end{abstract}

\section{Introduction}
\IEEEPARstart{T}{he} next-generation communication is expected to present revolutionary breakthroughs in global interconnectivity, intelligent services, and the ultimate communications experience. Leveraging the advancement of key technologies such as artificial intelligence (AI), reconfigurable intelligence surfaces (RIS), and integrated sensing and communication (ISAC), future communication is expected to deliver higher data rates, lower latency, higher energy efficiency, and massive connectivity capabilities \cite{aryan}. These improvements enable the network to support cutting-edge applications, including immersive XR experiences, holographic communications, and the artificial intelligence of things (AIoT), driving the evolution of the information society into a smart society.

Nevertheless, future communications still face transmission challenges in terms of spectrum efficiency and robustness. On the one hand, with the exponential growth of communication requirements, spectrum resources are becoming increasingly constrained. {The conventional construction of communication systems has mainly been rooted in Shannon's information theory, with the study of communications centered on enhancing the accuracy and efficiency of symbol transmission between the transmitter and receiver \cite{CognitiveSemanticCommunication}.} This focus has yielded significant advancements, but as communication technologies evolve, the theoretical Shannon limit is increasingly being approached \cite{shannon}. {This makes it difficult to further improve transmission spectrum efficiency. On the other hand, in the presence of sophisticated dynamic environments and diversified application requirements, the reliability and adaptability of conventional communication systems remain to be upgraded.} In this context, it is difficult to meet the requirements of future networks by relying only on the conventional communication paradigm. To address these limitations, semantic communication (SemCom) has emerged as a promising paradigm.

{In typical SemCom systems, conventional codecs are replaced by semantic codecs that leverage machine learning (ML) techniques.} The semantic encoder at the transmitter extracts the underlying meaning behind the transmitted information, rather than simply encoding raw symbols. It then aligns with the semantic knowledge base (SKB) and transmits only these extracted meanings. On the receiver, the semantic decoder with SKB reconstructs the intended message based on the received semantic representation \cite{qin}. By prioritizing the transmission of meaning rather than raw data, SemCom significantly reduces the volume of transmitted information, thereby improving spectral efficiency. Furthermore, SemCom enhances transmission robustness by reducing dependency on precise symbol-level accuracy, making the communication system more resilient to channel impairments and noise.

{Predictably, the integration of SemCom into future wireless networks has the potential to significantly enhance the quality of network services. Nevertheless, a key challenge in SemCom from the task-oriented nature of semantic codecs. As network conditions and application scenarios evolve, semantic drift occurs, wherein the meaning of transmitted information gradually degrades or diverges from its original interpretation, leading to a decline in communication accuracy. }This phenomenon necessitates continual updates to diverse modal (\emph{e.g.}, text and image) codecs to maintain efficient and reliable communication \cite{qin}. In end-to-end communication, such updates require the active participation of both the transmitter and the receiver to synchronize their semantic encoder and decoder. In the network, this challenge becomes even more complex. Unlike a simple two-party scenario, network-wide updates involve multiple users employing SemCom, each relying on a shared but evolving semantic understanding. Coordinating these updates across numerous parties significantly increases the complexity of maintaining a consistent and up-to-date semantic encoder-decoder model throughout the network. Managing these updates efficiently is crucial for ensuring seamless communication and maintaining the robustness of SemCom-enabled networks.

Several studies have focused on the challenge of collaborative semantic codec update among multi-parties and identified it as one of the primary challenges for SemCom applications \cite{commag}. Distributed learning approaches look most promising for user collaboration on codec updates as the privacy of the data required for user updates can be preserved. The classical federated learning (FL) approach for semantic codec update is first proposed for general networks to prevent data privacy \cite{5,6}. According to the different network scenarios' limitations, various schemes based on FL improvement are also presented \cite{uav,tgcn,iotj}.

\begin{figure*} 
\centering
\includegraphics[width=0.69\textwidth]{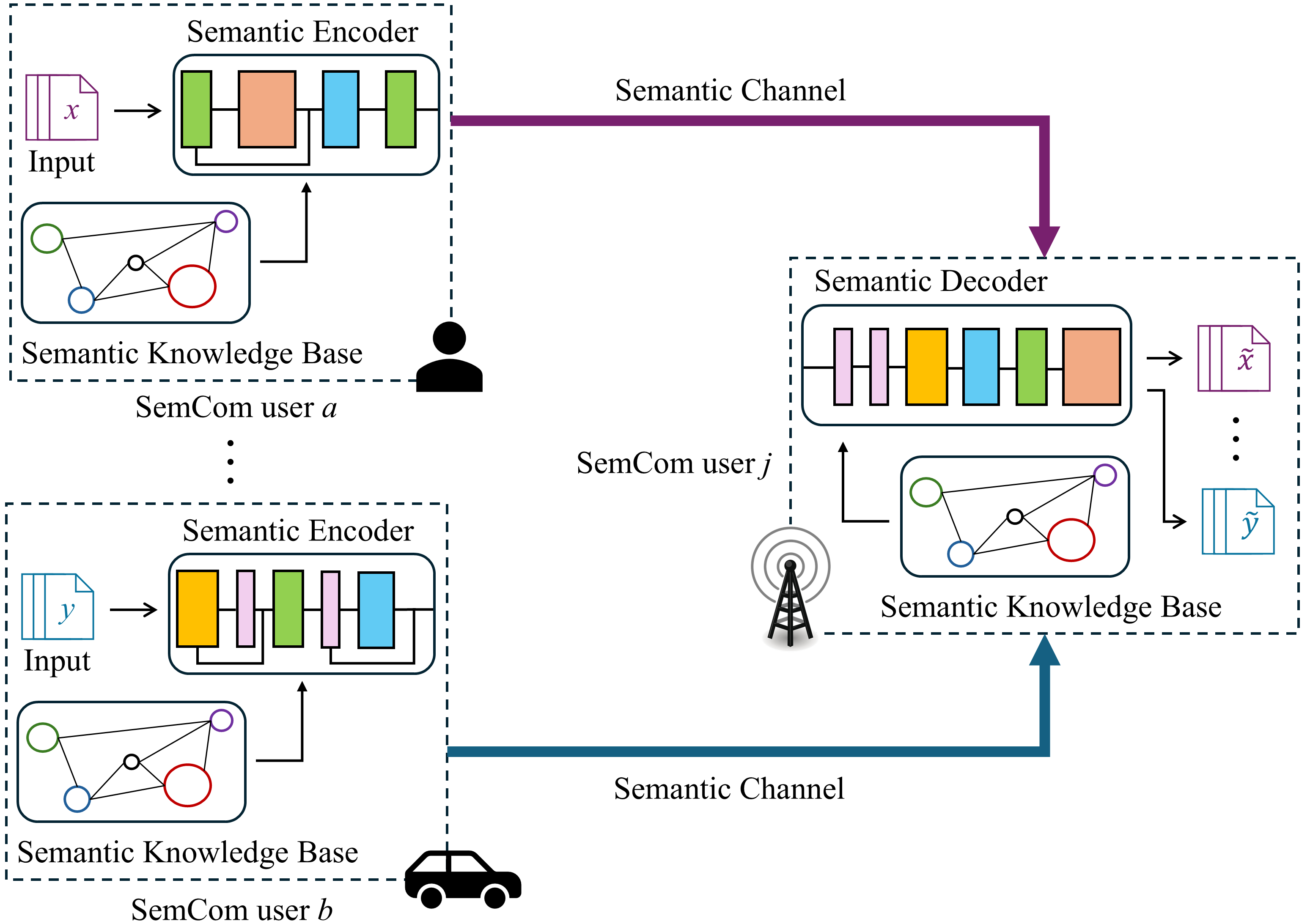} 
\caption{The SemCom framework.} 
\end{figure*}

In addition, growing attention is being given to the heterogeneous characteristics of network users during the semantic codec updating process \cite{commag,heto1,heto2}. {This consideration is particularly crucial in diverse network environments, such as non-terrestrial networks and vehicular networks, where significant heterogeneity exists in terms of users' (\emph{e.g.}, autonomous vehicles and small sensors) codec architectures, computational capabilities, etc.} In case the network environment undergoes changes, failure to perform joint updates across these heterogeneous components can lead to significant degradation in spectral efficiency. Users who do not adopt the newly updated semantic codecs may continue to operate under outdated codecs or a conventional communication paradigm, thereby creating bottlenecks in transmission accuracy and resource utilization, diminishing overall network performance. Hence, it becomes imperative to investigate adaptive semantic codec updating mechanisms in heterogeneous environments. It aims to ensure seamless compatibility and optimal functionality across evolving networks, enhancing both interoperability and spectral efficiency in SemCom systems.

Existing methods designed for general machine learning (ML) models, however, are not always directly applicable to semantic codec updating in heterogeneous networks. Unlike conventional ML-based models, semantic codecs must ensure synchronization and compatibility between two communicating parties, \emph{i.e.}, the transmitter and the receiver, to maintain effective task-oriented communication. This dual-party dependency introduces unique challenges, including the need for real-time coordination, adaptive learning mechanisms, and robustness against heterogeneous network conditions. Addressing these unique concerns is essential for ensuring seamless and reliable semantic communication across diverse and evolving network environments.

Despite the above-mentioned collaborative semantic codec update efforts, the study and development of semantic codec updates in heterogeneous networks are still at an early stage. Many technical challenges still need to be addressed to enable SemCom efficient and scalable deployment in real-world networks. Therefore, it is crucial to investigate these challenges associated with semantic codec updates in heterogeneous networks to help pave the way for more robust, efficient, and scalable solutions and facilitate the widespread adoption of SemCom across diverse communication scenarios.

In this article, we first review the key components in classical SemCom systems. We then discuss the main challenges of semantic codec updating in the heterogeneous network and potential methods. Next, we propose a heterogeneity-aware semantic codec updating scheme as a viable solution. The open issues are also discussed following that.

\section{Basic Networking Architecture}
As shown in Fig. 1, we review classical ML-based SemCom components in heterogeneous networks.

\subsection{Semantic Knowledge Base}

The SKB serves as a centralized repository of background knowledge and contextual information pertinent to the ongoing SemCom task. It contains task-related ontologies, domain-specific concepts, and relationships among entities that enable the semantic encoder and decoder to operate beyond symbol-level processing. By utilizing the SKB, the encoder can map raw data into semantically meaningful representations aligned with the intended task, while the decoder can accurately reconstruct and interpret the conveyed meaning even under channel distortions or incomplete data. Moreover, the SKB is shared between the transmitter and receiver, ensuring that both parties rely on an identical semantic context during encoding and decoding. This shared foundation enhances mutual understanding, reduces ambiguity, and improves the overall efficiency and robustness of SemCom across network conditions.

\subsection{Semantic Encoder and Decoder}
The semantic encoder, residing at the transmitter, is responsible for extracting and abstracting task-relevant meaning from the input data. Different from conventional communication paradigms, it employs ML to transform raw different modal inputs into a semantic representation that preserves the underlying intent and contextual meaning.

The semantic decoder, located at the receiver, reconstructs the conveyed semantics from the received representation. Accurate interpretation depends on close collaboration with the encoder, as even minor drift between the two can introduce semantic errors or task-level misinterpretations. Hence, to maintain fidelity across heterogeneous network conditions, the encoder and decoder for various modalities must be collaboratively updated in real time, ensuring that both parties share an aligned and adaptive semantic codec.

\subsection{Semantic Channel}
The semantic channel is a medium for information transmission, similar to the physical channel in traditional communication. The concern of the semantic channel, however, is not only the transmission of signals but also semantic integrity and fidelity. In practical communications, this may involve noise in the physical channel as well as noise at the semantic level (\emph{e.g.}, misinterpretation of semantics at the receiver). The robustness of the transmission is usually better than in conventional communications due to the small quantity of data and the fault tolerance of the semantic information.

\section{Key Challenges}
{The key challenges in updating the semantic codec model in heterogeneous networks include system heterogeneity, data heterogeneity, model heterogeneity, and personalized many-to-one (M2O) model requirements.}

\subsection{System Heterogeneity}
In a heterogeneous network, diverse users, \emph{e.g.}, autonomous-driving vehicles, portable carrying devices, and edge clouds, transmit information through SemCom under markedly different operating conditions. They vary in computational capabilities (\emph{e.g.}, CPU cycles) and communication transmission capabilities (\emph{e.g.}, transmission power, bandwidth). Furthermore, their communication capabilities are affected by dynamic channel conditions and noise. These disparities create pronounced system heterogeneity, making it challenging to achieve mutual understanding and coordination during collaborative updating.

\subsection{Data Heterogeneity}
Since semantic codecs are learned models, users locally update them using private data, which inevitably introduces data heterogeneity in terms of data size, distribution, modality, and semantic granularity. Data heterogeneity in semantic communication is task-oriented and semantic level. Users may observe different modalities (e.g., text, image, and speech) or different semantic contexts, which directly impact semantic alignment and meaning reconstruction between the transmitter and receiver. More importantly, this effect manifests over time during online communication. Heterogeneous local data can gradually induce inconsistent semantic mappings across users, leading to semantic drift even when local training converges, and ultimately causing task-level misinterpretation or communication failure.

\subsection{Model Heterogeneity}
System and data heterogeneity naturally lead users to adopt different semantic codec architectures with varying model sizes, computational complexity, and modality-specific designs. For instance, some users may employ a large pre-trained encoder optimized for high transmission accuracy for one modality. Others may rely on a smaller, lightweight model designed for faster computation with lower resource consumption for another modality. In the case of combining various modalities of data, updating multiple modality-specific codecs substantially increases the complexity and energy cost of the updating process. This disparity makes it difficult to establish a generally applicable global semantic codec, as users have varying constraints related to storage capacity, computational power, pre-training cost, semantic granularity requirements, and modality-specific needs. Users tend to use their initial encoder/decoder architectures, which render general collaborative training approaches inapplicable, \emph{e.g.}, FL, as they typically require model homogeneity for aggregation. Model heterogeneity in SemCom is further constrained by the tight coupling between the encoder and decoder deployed at different communication parties. Heterogeneous semantic encoders and decoders must remain mutually interpretable to ensure consistent semantic understanding during transmission. Over time, independent updates of heterogeneous models can gradually distort the shared semantic space, leading to incompatibility between transmitters and receivers even if each local model is individually well optimized.

\subsection{Personalized Many-to-One Model Requirements}

In case users collaborate on updating semantic codecs, it is impractical for one user to tailor the appropriate, personalized encoders or decoders for all other users. We take the downlink as an example. The base station utilizes the semantic encoder to transmit the information to personal devices equipped with semantic decoders. From a cost-efficiency perspective, maintaining and updating separate semantic encoders for a large number of personal devices imposes a substantial computational burden on the base station. Moreover, storing multiple distinct encoders significantly increases storage overhead, making such an approach infeasible in large-scale networks. Base stations thus aim to deploy a single, efficient semantic encoder capable of extracting and transmitting information that can be interpreted by various personal devices. Each personal device, in turn, utilizes a personalized heterogeneous semantic decoder to reconstruct the transmitted message based on its specific contextual understanding. This paradigm introduces a personalized M2O challenge, where a single (homogeneous) encoder/decoder needs to effectively serve a diverse (heterogeneous) set of decoders/encoders, each with unique semantic interpretations and requirements.

\section{Main Potential Methods}
In this section, we introduce several potential approaches to mitigate the challenges confronting SemCom in heterogeneous networks.
\subsection{Updating Based On Federated Learning}
FL-based methods enable collaborative semantic codec updating across distributed devices while preserving data privacy. By allowing users to locally update semantic codecs and share only model parameters, rather than raw data, FL is well-suited for heterogeneous networks where system heterogeneity and data heterogeneity. This approach can facilitate efficient semantic codec updates without centralized data collection, making it a promising solution for scalable and privacy-aware SemCom systems.

Nevertheless, the FL struggles with model heterogeneity and data heterogeneity, since different modal encoder/decoder architectures cannot be directly aggregated. {Even with homogeneous models, the averaged global model often fails to address personalized M2O model requirements, as a single shared model cannot optimally serve all receivers.}

\subsection{Updating Based On Split Learning}
Split learning (SL)-based approaches partition the updating process between distributed devices and a central server, enabling collaborative codec updating while preserving data privacy and reducing computational burdens on individual devices. In a typical setup, the transmitter updates the semantic encoder locally, while the receiver updates the corresponding decoder. Throughout the updating process, only intermediate features during updating are exchanged between the parties rather than the codec model, thereby sharing data while ensuring privacy. {It is thus particularly advantageous in personalized one-to-many model requirements, system heterogeneity, and model heterogeneity, as it allows distributed users to train their personalized encoder/decoder and a joint decoder/encoder while ensuring synchronization between heterogeneous components.}

Nonetheless, in order to achieve the Loss value, the privacy training data and labels need to be known by the encoder parties and the decoder parties, respectively. It introduces critical privacy vulnerabilities. Moreover, SL-based approaches perform poorly with data heterogeneity and data heterogeneity. The reliance on local data for loss calculation causes the modality-specific codec to overfit particular data distributions, especially under non-IID conditions. This bias is worsened by its sequential updating, which skews the global model toward the latest user's data.

\subsection{Updating Based On Transfer Learning}
Transfer Learning (TL) leverages pre-trained semantic codecs to adapt to new tasks or environments with minimal retraining. {By updating models using only local data without relying on collaborative aggregation, TL-based methods effectively mitigate model heterogeneity, system heterogeneity, and personalized M2O model requirements, etc.} For example, a base station could deploy a universal encoder, while users personalize their decoders via TL.

However, TL  involves fine-tuning only the latter parts of the model while keeping the earlier layers fixed, which limits its flexibility in certain scenarios. For example, a base station may deploy a universal decoder, while users struggle to effectively personalize their encoders through TL. Furthermore, the lack of model/data sharing results in missing data distributions during updating. Designing distinct transfer strategies is also necessary for data heterogeneity, increasing the complexity of model adaptation and storage load.

\subsection{Hybrid Approaches}
The integration of FL, SL, and TL has shown considerable promise in mitigating the individual limitations of each paradigm when applied independently. For instance, federated transfer learning has been effectively introduced into vehicular SemCom networks to improve knowledge sharing across structurally and functionally heterogeneous nodes \cite{commag}. Similarly, federated split learning frameworks have been proposed to enhance collaborative training while reducing computational burdens and communication overheads in distributed settings \cite{heto2, jsac}.

{However, while these methods have partially alleviated challenges related to heterogeneity, \emph{e.g.}, system heterogeneity \cite{jsac}, model heterogeneity \cite{heto2}, and personalized M2O model requirements \cite{commag}, they still fall short of fully accommodating the multifaceted updating requirements imposed by heterogeneous networks.} There remains a critical need for adaptive, scalable, and granular updating schemes that can respond to varying degrees of heterogeneity across devices, data, and modalities.

\section{Heterogeneity-aware Semantic Codec Updating Scheme}
{To cope with the heterogeneity of networks in future semantic communication systems, in this section, we propose a heterogeneity-aware semantic codec updating scheme.} We decouple encoder updates from decoder maintenance by aligning all modality-specific representations to a shared SKB, enabling lightweight and efficient updating via FL.

\begin{figure*} 
\centering
\includegraphics[width=0.69\textwidth]{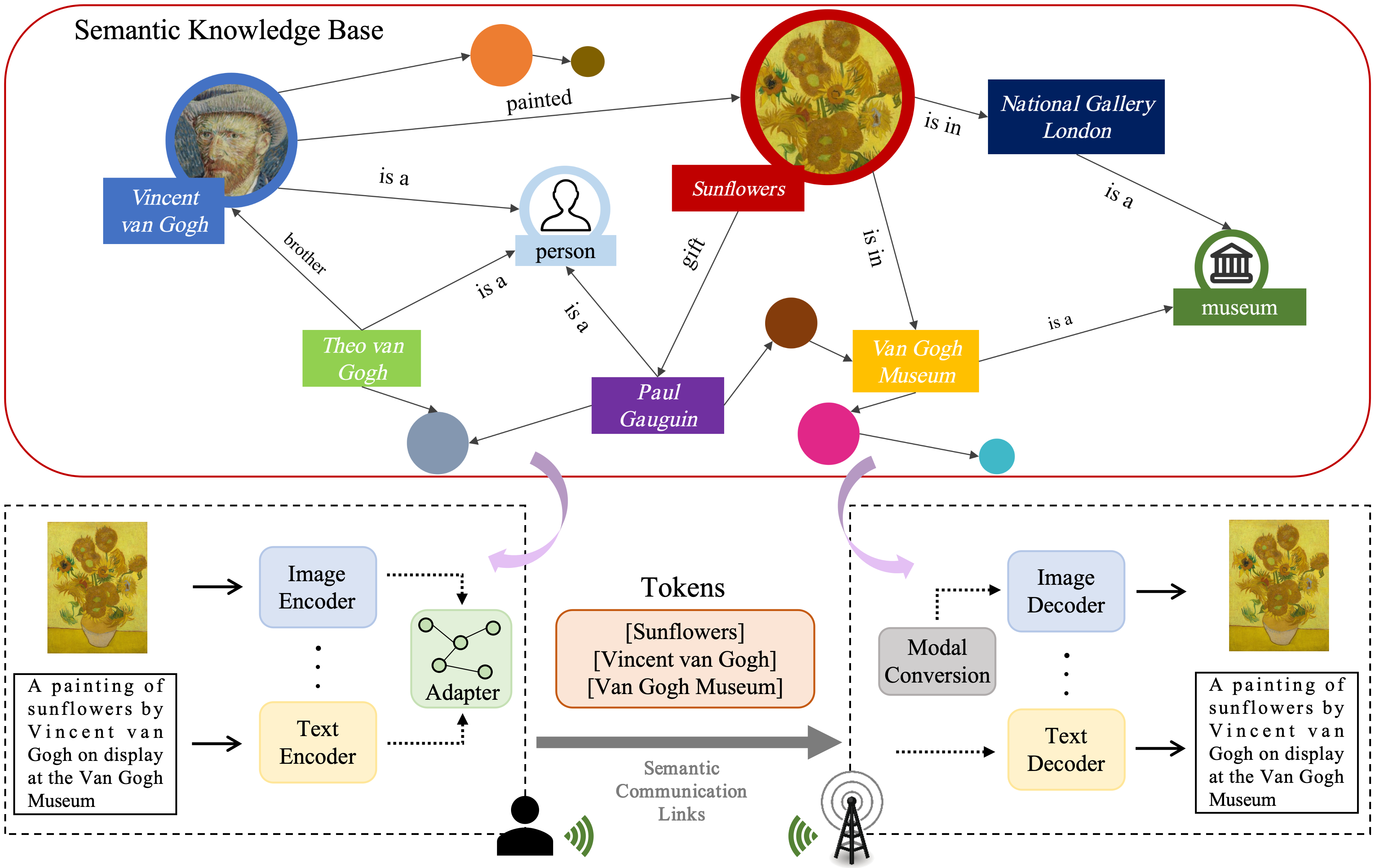} 
\caption{{SKB–assisted Modal Alignment.} }
\end{figure*}

\subsection{SKB–assisted Modal Alignment}
Instead of retraining the entire encoder–decoder pair for every user, the proposed scheme only collaboratively updates the encoder part by utilizing the centralized SKB that defines a unified semantic space. Each user retains a frozen heterogeneous encoder to extract modality-specific features, \emph{e.g.}, image, speech, or text. To make the encoder model federated aggregation available for modal and model heterogeneity, we designed a global adapter capable of integrating diverse modalities and encoder architectures. A global multi-modal adapter further performs semantic feature extraction and transforms these multimodal features into semantic tokens aligned with the SKB, and only the aligned tokens are transmitted. At the receiver, the decoder reconstructs the intended meaning by consulting the same SKB, effectively isolating the decoder from upstream drift. {By shifting the update focus entirely to the encoder side, in particular, the global adapter, this design mitigates M2O personalization issues, accommodates system and model heterogeneity, and streamlines semantic codec updating.}

\subsection{Adapter Construction}
The multimodal adapter is realized through a graph-based network that operates on a unified semantic graph built from heterogeneous modal inputs. Each modality output from the heterogeneous encoder is transformed into "tokens" for transmission, and each modality region or token becomes a node in the graph. The same modal relations are captured by Gaussian-kernel similarities with top-k sparsification to emphasize salient dependencies. Moreover, associations across modalities are derived by computing node similarity and refining it through a lightweight neural mapping with top-k selection and mean pooling, yielding edges that encode semantic correspondence across modalities. Through iterative message passing, the adapter propagates semantic cues, aligns disparate features with the dimensions of the SKB, and produces modality-invariant token embeddings ready for transmission. This graph-based design naturally scales to different modalities and allows fine-grained control over semantic interactions.

\subsection{Error-aware Labeling for Adapter Updating}
Nevertheless, in full encoder–decoder training, the model can directly compare input and reconstructed output to compute a well-defined loss \cite{jsac}. In our proposed scheme, the encoders are frozen and only update the global adapter, which references the target labels disappearing. Encoder-only adaptation, hence, disrupts the conventional SemCom codec updating loop because reliable ground-truth labels for aligned tokens are unavailable during live operation. To circumvent this, the adapter interprets its own SKB-aligned output as a provisional label. However, such labels inevitably contain semantic drift, \emph{i.e.}, most tokens remain correct, but a minority are mismatched. To maintain training stability, before each update epoch, the adapter performs an estimate of the similarity matrix across a mini-batch of multi-modal samples. Confidence for each label/token is inferred from the matrix, and low-confidence labels are temporarily excluded from encoder gradient updates in this epoch.

Federated aggregation across multiple users further improves label quality. Because different users observe distinct but related data, their error patterns are unlikely to coincide. Aggregating label statistics exposes outliers and reinforces consensus on correct alignments, effectively mitigating error without explicit ground truth. The users, thus, do not immediately discard low-confidence labels but track them. They will keep these potential labels across several rounds, and only those persistently inconsistent with other federated aggregation users are pruned. Multiple users collaborate through federated aggregation, cross-validating labels, and leveraging inter-user diversity to suppress error and recover accurate supervision. This dynamic denoising mechanism preserves semantic alignment even under local drift or highly non-IID data, ensuring that encoder updates remain robust while the decoder remains untouched.

By anchoring all representations to a shared SKB and confining updates to lightweight adapters, the proposed scheme reduces communication cost, preserves decoder fidelity, and enables scalable personalization. The graph-based adapter and label error-aware training jointly empower cross-modal generalization, making the approach a promising foundation for robust SemCom updating in heterogeneous networks.

\subsection{Case Study}
To evaluate the performance of the proposed heterogeneity-aware semantic codec updating scheme, we constructed a distributed network comprising 10 heterogeneous users. Each user transmits various modal data from the Flickr30K dataset \cite{sim} using either a Faster R-CNN image encoder or the BERT-base-uncased text encoder. The dataset contains 31,000 images and 158,000 captions, making it well-suited for semantic transmission tasks involving multiple modalities. We adopt RSUM as the evaluation metric, defined as the sum of accuracies in both image-to-text and text-to-image retrieval tasks performed by the adapter guided by the SKB. This metric reflects alignment with the SKB and indicates the precision of model updates. All simulations were conducted on the Lancaster University HEC GPU cluster using NVIDIA V100 (32 GB) and L40 (48 GB) GPUs via the gpu-short and gpu-medium queues. In terms of model configuration, the adapter consists of a single-layer graph neural network with an attention scaling factor set to 1.5, a regularization weight of 1.0, and a triplet loss margin of 0.2. During training, a batch size of 128 is used along with the Adam optimizer at a learning rate of 0.0001.

\begin{figure}[t] 
\centering
\includegraphics[width=0.5\textwidth]{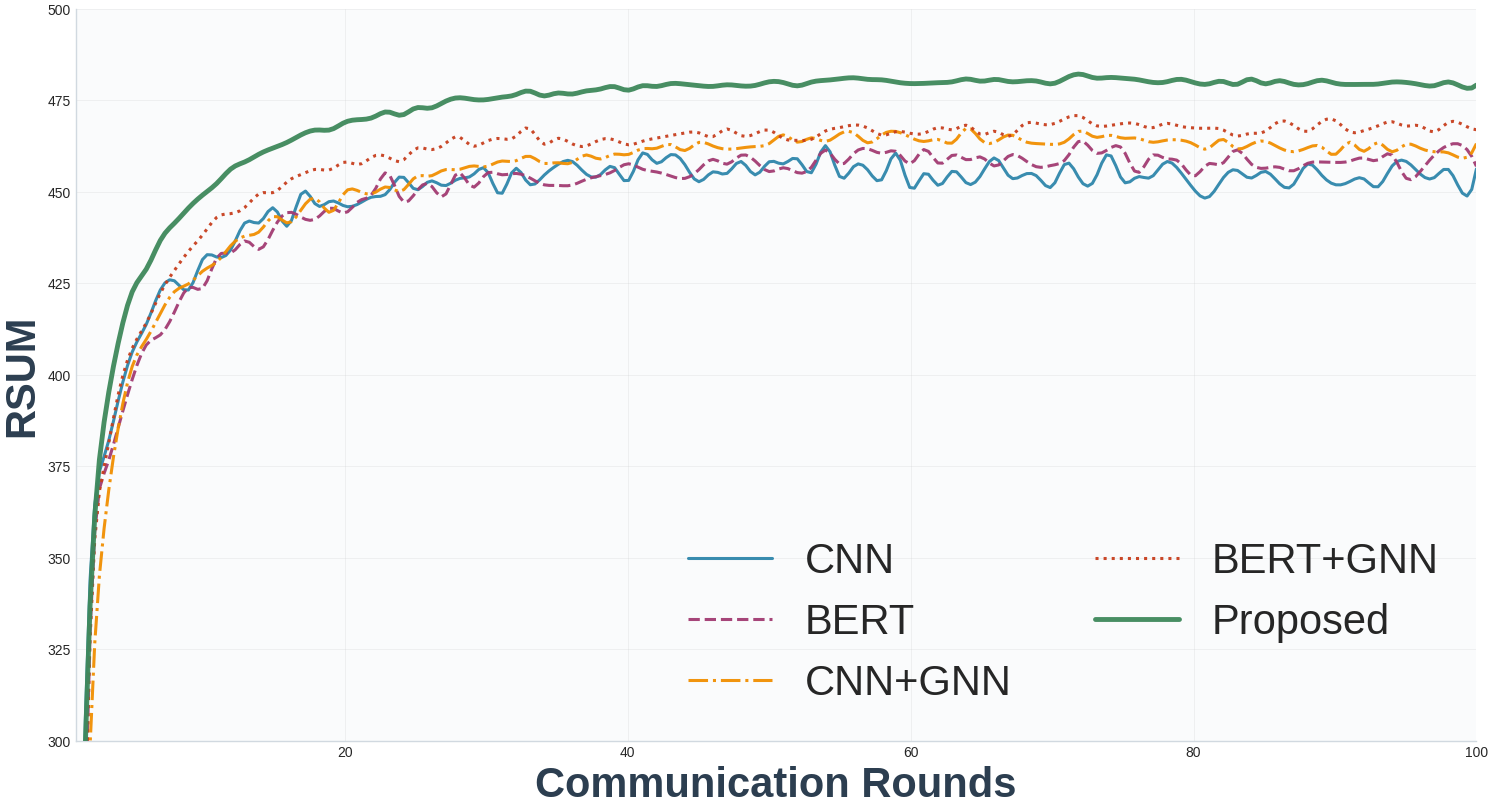} 
\caption{Subscriber access number versus number of subscribers.} 
\label{4}
\end{figure}

\begin{figure}[htbp]
    \centering
    
    \begin{subfigure}[b]{0.5\textwidth}
        \centering
        \includegraphics[width=\textwidth]{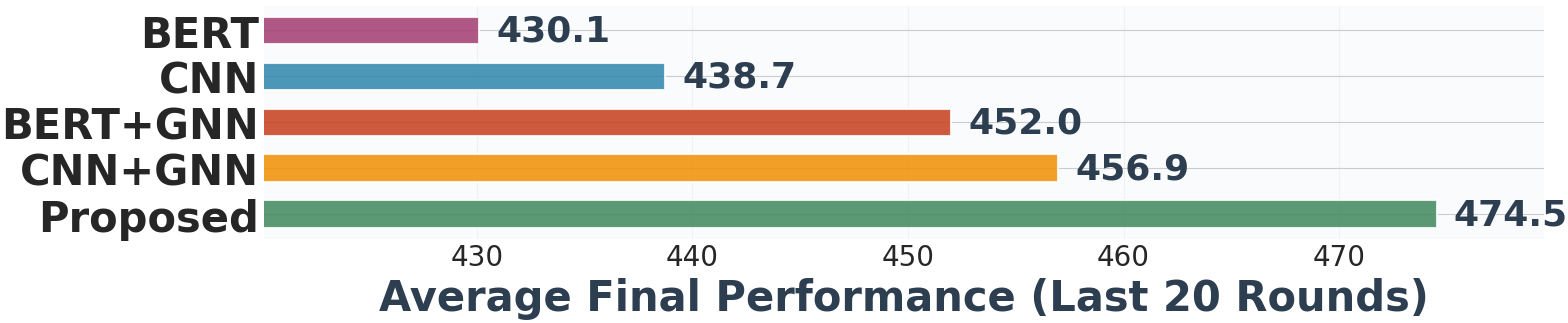}
        \caption{User with mild label error.}
        \label{fig:sub1}
    \end{subfigure}
    
    \vspace{0.5cm} 
    
    \begin{subfigure}[b]{0.5\textwidth}
        \centering
        \includegraphics[width=\textwidth]{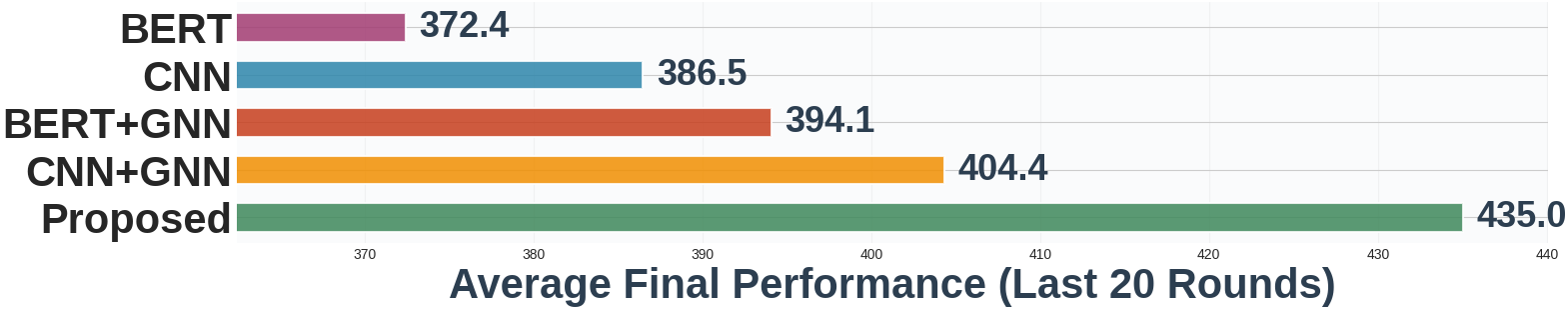}
        \caption{User with severe label error.}
        \label{fig:sub2}
    \end{subfigure}
    
    \caption{Performance comparison under varying levels of label error.}
    \label{fig:total}
\end{figure}

We first examine a baseline scenario where users perform updates using clean (error-free) labels in the heterogeneous network. Under this setting, users update their models directly with correct supervision. To handle heterogeneity, each user employs a lightweight CNN or BERT model with a uniform architecture for single-modal federated aggregation \cite{6} for baseline. Additionally, a GNN model \cite{end} serves as a cross-modal adapter to aggregate updates from heterogeneous client models. As illustrated in Figure 3, the proposed approach surpasses conventional methods in both convergence speed and accuracy, confirming the efficacy of the proposed updating scheme in achieving precise model evolution.

We further investigate a more challenging scenario where real-time on-device training introduces label error due to semantic drift. To mitigate the effects of randomness, we divide users into two groups of five users each, \emph{i.e.}, those affected by mild error and those affected by severe error. Fig. 4(a) and (b) compare the accuracy of users under various training schemes under different error levels. The proposed method consistently achieves the best final accuracy in both groups, particularly in severely affected environments. These results demonstrate its robustness and practical relevance in real-world scenarios with noisy label conditions. Moreover, the proposed adapter-based updating scheme significantly reduces both computational and communication overhead compared to traditional full-model updates, as only the lightweight adapter parameters are trained and transmitted, while the larger encoders remain frozen.

\section{Open Research Topics}
In previous sections, we have shown that the SemCom is expected to be adaptive updating in heterogeneous networks. However, this research is still in its infancy. This section discusses open research problems and possible solutions in heterogeneous SemCom.

\subsection{Discrimination Arising from Heterogeneous Model and System}
Due to the heterogeneity of the semantic codec models, the traditional model aggregation approaches are not applicable to the collaborative updating of semantic codecs. Leaving heterogeneous models for local training causes incomplete information between users. Each user does not know each other's data information, model information, and system information at the time of training. In the case of one user being slow to update due to model or system reasons, other users may exclude that user from joining the update due to considerations such as accuracy or update efficiency. This occurs even if the excluded user possesses richer data and could achieve higher transmission accuracy after updating, since others lack this information. This prevents this user from deploying semantic communication properly due to discrimination during updating, thus necessitating a high communication load via employing conventional communication paradigm. To mitigate this, potential solutions such as developing social-inspired anti-discrimination mechanisms that tolerate stragglers could be explored.

\subsection{Fairness issues with Heterogeneous Data}
Datasets of users are generally considered non-IID data. Yet, inconsistencies in the data ‘class’ not only influence updates to transmission accuracy and convergence, but also lead to an unfair SemCom after updating, since the user only gets their personalized data and cannot determine the others’ data. For instance, in the United Kingdom, a minority ethnic user performs collaborative semantic codec updates with other users in a community of Britons. His training information is related to his ethnicity, \emph{e.g.}, pictures and language. This information, however, is a minority in this community, making the updated semantic codec less accurate for him than for the local Briton community users. This creates an unfair update for the minority ethnic user, who is also involved in the update. Potential solutions include the adoption of fairness-aware aggregation algorithms or the introduction of gradient correction mechanisms to mitigate bias against underrepresented data distributions.

\subsection{{Privacy Risk of Personalized M2O Model}}
{In the network, there is an M2O characteristic of semantic models, the semantic encoders and decoders that should form an integral ML model are distributed across different users.} During the updating of the SemCom model, if a malicious user can get hold of the semantic encoder/decoder model of one normal user, it can not only affect the coding/decoding of that user but also cause the decoding/coding methods of other users to be deduced. This leads to leakage of transmitted information after the SemCom model updating is complete, even though each user's semantic model is individual and heterogeneous. Hence, the semantic codec model is related to the accuracy of signal transmission. Users need to consider not only transmission data privacy and system information privacy, but also, crucially, to consider the semantic codec model privacy while employing SemCom. To enhance privacy, methods such as homomorphic encryption for model updates or differential privacy in collaborative learning could be explored.

\subsection{M2O Collaborate SemCom/Inference}

In the M2O SemCom paradigm, network heterogeneities critically degrade communication reliability and accuracy. It also influences the SemCom derivative paradigm, collaborative inference (CI), in which multiple devices jointly infer via semantics toward a common result. For instance, data heterogeneity challenges the fusion of disparate inputs into a coherent semantic representation, model heterogeneity introduces inconsistent feature abstractions that obscure intended meanings, and system heterogeneity corrupts transmitted semantics through varying channel conditions. These compounded effects cause the final SemCom/CI outcome to substantially deviate from the ideal, highlighting the necessity for robustness-aware M2O SemCom designs. Potential solutions could involve a graph-based semantic adapter to unify heterogeneous features.

\subsection{Dynamics in Heterogeneous Network}

Several studies have discussed the integration of semantic communication into practical networks, which are often dynamic, \emph{e.g.}, vehicular networks\cite{iotj} and satellite networks\cite{jsac}. In such networks for semantic codec updating, there exist new users joining and veteran users leaving. The locations of users are also changed. It is thus urgently needed to address the above issues, such as privacy in heterogeneous networks, while taking into account the dynamics of the user during semantic encoder updates. { Potential solutions involve designing lightweight onboarding protocols for new users and elastic aggregation strategies that adapt to network churn.}

\subsection{Incentive Mechanisms for Semantic Codec Updates}
Facing a brand new task, users in the network associate to perform semantic codec updates to transmit spectral efficiency as well as robustness enhancements for that task. Furthermore, due to heterogeneity, each user is not equally enthusiastic in the same sense as to whether or not to transmit for this task using SemCom, \emph{i.e.}, whether or not to participate in the update. Nevertheless, if some users participate in the update and some refuse, it does not simply affect a single user. Because communication is mutual, users of two different communication paradigms cannot communicate. Moreover, network spectrum resources are limited, and users who refuse to update. To encourage participation, incentive mechanisms such as updated credit rewards or resource allocation priority could be introduced.

\section{Conclusions}
This article surveyed the deployment of SemCom in heterogeneous networks, highlighting the challenges during updating semantic codecs to facilitate the widespread adoption of SemCom. Moreover, a heterogeneity-aware semantic codec updating scheme was proposed and constitutes a promising candidate for supporting SemCom in heterogeneous networks. Lastly, we also discussed some open issues, such as discrimination, fairness, privacy, network dynamics, and updating enthusiasm. We hope that the challenges and opportunities in this article will pave the way to advance SemCom for large-scale applications on the network in the future.

\section{Acknowledgments}
This work was supported by the National Natural Science Foundation of China under Grant 62571307, the National Key Research and Development Program of China under Grants 2022YFB2902304, the Science and Technology Commission Foundation of Shanghai under Grants 24DP1500500, and the Open Networks Ecosystem Competition Western Open Radio Access Network (O-RAN) Deployment (ONE WORD) Project. The authors would like to give special thanks to the Van Gogh Museum, Amsterdam, for granting permission to reproduce the images of Vincent van Gogh’s Sunflowers (s0031V1962) and Self-Portrait with Grey Felt Hat (s0016V1962) in this article.

\ifCLASSOPTIONcaptionsoff
  \newpage
\fi

\textbf{Guhan Zheng} is an Associate Professor at the School of Communication and Information Engineering, Shanghai University, Shanghai, China. His research interests include non-terrestrial networks, semantic communication, and network economics.

\textbf{Qiang Ni} is a Professor at the School of Computing and Communications, Lancaster University, Lancaster, U.K. His research areas include future generation communications and networking, including green communications/networking, millimeter-wave wireless, cognitive radio systems, 5G/6G, SDN, cloud networks, edge computing, dispersed computing, Internet of Things, cyber physical systems, artificial intelligence/machine learning, and vehicular networks.

\textbf{Aryan Kaushik} is the Chief Innovation Officer (CIO) at RakFort, Ireland, since 2025. He has been an Associate Professor at Manchester Met, UK, from 2024-25, and an Assistant Professor (Senior Grade) at the University of Sussex, UK (2021-24). His research interests include signal processing for 5G/6G wireless communications, integrated sensing and communications, reconfigurable holographic and intelligent surfaces, energy efficient communication systems, non-terrestrial networks (satellite and UAV Communications), millimeter wave massive MIMO, and machine learning, edge computing, and AI-assisted communications.

\textbf{Lixia Yang} is a Professor at the School of Electronic and Information, Anhui University, Hefei, China. His current research interests include signal analysis and design, electromagnetic wave propagation and scattering in complex media and objects, computational electromagnetics, and inverse scattering.

\textbf{Yushi Wang} is a Ph.D. candidate at the School of Computing and Communications, Lancaster University, Lancaster, U.K. Her research interests include federated learning and semantic communications.

\textbf{Charilaos Zarakovitis} is a Research Director at the ICT Department, Axon Logic, Greece. His research interests include machine and deep learning, green communications modelling, bioinspired and game-theoretic decision-making, cognitive radios, network virtualisation, quantum neural networks, statistical signal processing, and convex optimisation analysis.

\end{document}